\documentclass[twocolumn,prc,aps,floatfix]{revtex4}
\usepackage[dvips]{graphicx}
\begin{document}
\title{Exchange-assisted tunneling in the classical limit}
\author{V.V. Flambaum}
\affiliation{
 School of Physics, The University of New South Wales, Sydney NSW
2052, Australia
}
\date{\today}
\begin{abstract}
The exchange interaction and correlations may produce
 a power-law decay
 instead of the usual exponential decrease of the wave function under
 potential barrier.
The  exchange-assisted tunneling vanishes in the classical limit, however,
the dependence on the Planck constant 
$\hbar$ is  different from that for a conventional
 single-particle tunneling.
\end{abstract}
\maketitle

\section{Introduction}
In the Hartree-Fock equations the exchange interaction is described by
 the non-local (integration) operator  $ K({\bf r})$, and the well-known
 theorems proven for the Schroedinger equation with a local potential
 $U({\bf r})$ are violated if we add the exchange term (or any other
 non-local operator). The Hartree-Fock equation for a fermion
 (e.g. bound atomic electron)
 at orbital  $\Psi_b({\bf r})$ has the following form:
\begin{equation}\label{HF}
-\frac{\hbar ^2}{2m}\frac{d^2}{d{\bf r}^2} \Psi_b({\bf r})
 +(U({\bf r})-E) \Psi_b({\bf r})= K({\bf r})
\end{equation}
\begin{equation}\label{HFK}
 K({\bf r})=\sum_q \Psi_q({\bf r}) \int \Psi_q({\bf r'})^\dagger \frac{e^2}{|{\bf r-r'}|} \Psi_b({\bf r'}) d {\bf r'}
\end{equation}
Here the summation runs over all occupied orbitals  $\Psi_q({\bf r})$ with
 the same spin projection as  $\Psi({\bf r})$. The range of $ K({\bf r})$
is determined by the range of the highest orbitals $\Psi_q({\bf r})$ which
stay outside the integral in Eq. (\ref{HFK}). If  $\Psi_q({\bf r})$ belongs
to continuum (like an electron state in a conducting band in a crystal), the
range is infinite. In this case the range of the bound electron solution
of  Eq. (\ref{HF}) (e.g. atomic 1s orbital) is also infinite since 
$\Psi_b({\bf r})$ can not  vanish if the right-hand-side of  Eq. (\ref{HF})
is not zero.

According to \cite{DFS,Flambaum2009}
 (see also \cite{FlambaumModel,Froese,Handy})
the exchange can produce a power-law decay instead of the
usual exponential decrease in a classically forbidden region.
 For inner orbitals inside
 molecules decay is $r^{-2}$ \cite{DFS}, for macroscopic systems
 $\cos{(k_f r)} r^{-\nu}$, where $k_f$ is the Fermi momentum and 
   $\nu=3$ for 1D, $\nu=$3.5 for 2D and $\nu=$4 for 3D crystal
 \cite{DFS,Flambaum2009}.
Correlation corrections within the perturbation theory approach
 do not change these conclusions
 \cite{Flambaum2009,FlambaumModel}.
 Slow decay increases the spin-spin interaction
 between localized spins in solids and the under-barrier tunneling amplitudes. 
According to Ref. \cite{Amusia} the exchange interaction
may increase probability of ionization of inner atomic electrons by an
 external electric field (e.g. laser field) by many orders of magnitude
 (a different enhancement mechanism,
an ``atomic antenna'', was suggested by Kuchiev \cite{Kuchiev}
and rediscovered by Corkum \cite{Corkum}).

In this paper we want to consider the problem in the
 classical limit $\hbar \to 0$.  In the case
of a single-particle problem a wave function under the potential
barrier is exponentially small, $\Psi \sim \exp{(-\int |p|dr/\hbar)}$,
i.e. the probability of the tunneling tends to the classical result
 (zero) very fast, it contains $\exp{(-1/\hbar)}$.
Here $|p|=\sqrt{2m(U(r)-E)}$ is the semiclassical under-barrier momentum.
 At first glance, one may say that the exchange
interaction does not exist in the classical physics, therefore, the
 probability of the exchange-assisted tunneling should also rapidly vanish.
Moreover, any effects of identity of particles (Fermi and Bose statistics)
 should vanish when the wavelength $\lambda=h/p$ tends to zero.
However, the answer may be  not as simple as it  looks.
For example, let us consider a  three-body problem.\\
A homogeneous electric field is applied to a hydrogen atom where the
 electron energy is $E_b$. If the field
is not too strong, the probability of the tunneling to the state
 of the same energy $E_b$ in  continuum outside the atom is practically zero.\\
Then another electron with energy $E_c$ (above the barrier) hits the atomic
electron, loses energy $E_c-E_b$  and occupies the state with energy
 $E_b$ (equal to the energy of the bound electron) in
 continuum outside the atom.\\
The lost energy $E_c-E_b$ is transfered to the  bound electron
 which gets enough energy
($E_c$) to pass the barrier. Both electrons are free and have the same
energies $E_b$ and $E_c$.\\
 This process imitates the ``tunneling''
of a particle with the energy  $E_b$ (for non-identical particles one more
 collision with the reverse energy exchange is needed to imitate the
 ``tunneling''). This picture looks like a purely classical one.
Therefore, it is not obvious why the exchange-assisted tunneling must
vanish  for $\hbar \to 0$. A related question: if this effect vanishes,
is the law exponential (similar to the vanishing of the  single-particle
 tunneling), or does it vanish less rapidly?

\section{Two-well potential}
To answer these questions it is instructive to consider a simple
 model with a minimal number of discrete states involved (to avoid
interference of different contributions during summation
 and integration over spectrum which may change the law of $\hbar$ dependence;
 this happens for the dependence on distance $r$ \cite{Flambaum2009}).

We start from  a  model of resonance tunneling from one potential
well to another potential well. The case of symmetric double-well potential
has been solved e.g. in the textbook \cite{Landau}. There are two levels
corresponding to the symmetric (ground state) and antisymmetric wave
 functions. The tunneling produces the splitting of these levels,
 $E_{\pm}=E_1 \mp t_1$ where
 $t_1 \sim \exp{(-\int |p|dr/\hbar)}$ is the tunneling amplitude,
and the integral is taken between the classical turning points.

 If the first  potential well  (``L'') is slightly deeper than the second
potential well (``R''),  the ground state wave
 function may be presented as $\psi_g=\psi_{1L}+B_{t1} \psi_{1R}$ where
$B_{t1} \sim t_1/(E_{1L}-E_{1R})$.
Here we assume that the distance to other levels is large,
 $t_1 \ll (E_{1L}-E_{1R})$ and the probability of the
 particle in the ground state to be in the well $R$ is exponentially small
(proportional to the squared tunneling amplitude,
 $B_{t1}^2 \sim t_1^2/(E_{1L}-E_{1R})^2$).

 Now we add a second particle  (identical fermion or boson) to a higher state 2
 which has energy
close to the top of the barrier.  We can present its wave function as
$\psi_2=A_2\psi_{2L}+B_2\psi_{2R}$ where the coefficient $B_2$ is not
 necessarily small.
In this case the probability
 of the particle in the ground orbital to be in the potential well $R$ is no
 longer proportional  to the exponentially small parameter $t_1^2$. Indeed,
the following two-step process takes place.

 Step 1: the second particle
  tunnels from the potential
well $L$ (orbital $\psi_{2L}$) to the potential well $R$ (orbital $\psi_{2R}$).

Step 2: two-body process $2R,1L \to 1R,2L$ due to  a nondiagonal Coulomb
 exchange interaction  which transfers the first  particle  from orbital
1$L$ to the orbital $2L$ and the second particle  from $2R$ to  $1R$.

 As a result of these two steps, we have no change
in the occupation of the state 2 and transfer of  a particle from the ground
 state $1L$ to $1R$.
This gives the amplitude for the ground state particle to be in the well 
``R'':
\begin{equation}\label{BG}
 B_{G1} \sim \frac{G(2,1L;1R,2)}{E_{1L}-E_{1R}},
\end{equation}
 where 
\begin{eqnarray}\label{G}
G(2,1L;1R,2) =\\
\nonumber 
\int \psi_{2}({\bf r})^\dagger\psi_{1L}({\bf r})
 \frac{e^2}{|{\bf r-r'}|} \psi_{1R}({\bf r'})^\dagger \psi_2({\bf r'})
 d {\bf r'}d {\bf r}
\end{eqnarray}
is the Coulomb exchange integral. Note that the potential wells here may
have one, two or three dimensions.

This result may also be derived from the Hartree-Fock equations
 (\ref{HF},\ref{HFK}) for the orbital
 $ \Psi_{b}= \psi_{1}= \psi_{1L}+ \delta \psi_{1}$,
by  projecting it to the orbital $ \psi_{1R}$. Note that the contribution of
 the direct term in the Coulomb interaction between the particles 1 and 2 is
 included into the mean field  potential $U({\bf r})$.

 Equation (\ref{G}) gives us dependence of the amplitude 
$B_{G1}$ on the distance $l$ between the wells $L$ and $R$.
If the distance $l \gg r_1$ where $r_1$ is the size of the orbital $1L$,
we can expand  $1/|{\bf r-r'}|$ near $|{\bf r-r'}|=l$.
Integral with the first term $1/l$ of this expansion vanishes due to the
 orthogonality of the wave functions $\psi_{1}({\bf r})$ and
 $\psi_{2}({\bf r})$. Therefore, the expansion starts from $1/l^2$.
The correlation effects correspond
to higher orders in the perturbation theory in the  Coulomb  interaction
 integrals G, so they decay with distance faster than  $1/l^2$.  

\section{Classical limit} 
Now consider the classical limit $\hbar \to 0$.
The exchange
term in the Hartree-Fock equations (\ref{HF},\ref{HFK})  seems to extend to a
large distance (under the barrier) if we consider the exchange interaction
 of an under-barrier
 particle ($\Psi_b=\psi_1$) with an above-barrier one ($\Psi_q=\psi_2$).
 The second term
(the first non-vanishing term) in the multipole expansion of $K(r)$
gives us $\psi_2(r)/r^2$ decay in the under-barrier area in the 
right-hand side  of Eq. (\ref{HF}). Contrary to the single-particle
tunneling amplitude  ($t_1 \sim \exp{(-\int |p|dr/\hbar)}$)
 this dependence does not contain $\hbar$.  How the classical limit may be
 reached in this case? The $\hbar$ dependence actually comes from the
coefficient before $\psi_2(r)/r^2$ which is given by  the integral in $K(r)$.

  The answer here depends on what kind of states we consider.\\
1. Two semi-classical states, one below the barrier, another above the barrier.
In this case we have integral between two rapidly oscillating functions in
 $K(r)$ in Eq.(\ref{HFK}),\\
 $\sim \int ({\bf dr'}/|{\bf r-r'}|)\cos{(\int p_1(r)|dr/\hbar)}
 \cos{(\int p_2(r') dr'/\hbar)}$.\\ 
The integrals of this type are considered in the textbook \cite{Landau}
in the chapter about matrix elements between semiclassical states.
Such integrals are exponentially small  at $\hbar \to 0$ 
 (they contain $\exp{(-1/\hbar)}$), therefore,
the coefficient before $\psi_2(r)$  in
 $K(r)$ in Eq.(\ref{HFK}) in  the right-hand-side
 of Eq. (\ref{HF}) (and the exchange- induced $\delta \psi_1$) vanishes.
\\  
2. Wave function of the first particle, $\psi_1$, is essentially a quantum
 state, a ground state or a low energy state. Here the result depends
 on the properties of the potential $U(r)$. If this potential
is smooth near the minimum, we may use the oscillator wave functions
to describe the low-energy states. The exchange integral in this case
contains  product of the oscillator function  and the
rapidly oscillating semiclassical wave function,
$\sim \exp{(-\xi^2/2)} \exp{(ipx/\hbar)}=
\exp{(-(\xi-\xi_0)^2/2)}\exp{(-p^2/(2m\omega \hbar))}$,\\
 where $\xi=x\sqrt{m\omega /\hbar}$. The exchange integral is
 exponentially small again (see the last exponent).\\ 
3. A singular attractive potential ( a model of a diatomic molecule).
In  this case the  size of the lowest states so rapidly tends to zero
  at $\hbar \to 0$ that there are  no oscillations inside.
  Consider, for example, a Coulomb ground 
state where the size is given by the Bohr radius $a_B=\hbar^2/(me^2)$.
At $r < a_B$ the product $(p/\hbar) r < 1$,
 i.e. there are no oscillations.
 Therefore, $K(r)$ has no exponential suppression for
 $\hbar \to 0$.  However, the size
 of the ground state ($\sim a_B$) tends to zero and
 the binding energy($\sim e^2/a_B$) to infinity,
 so this case actually has no classical limit.

\section{Conclusion}
We see that
 the exchange-assisted tunneling vanishes in the classical limit
(it still may be exponentially enhanced in comparison with the single-particle
tunneling).  The difference with the classical
``ionization by collision'' case is that in the latter we do not fix
final energy of each particle (provided the total final energy is equal
to the initial energy, $E_{1f}+E_{2f}=E_{1i}+E_{2i}$),
 the energy exchange depends
on the initial conditions. In the exchange-assisted
tunneling the spectrum of particle energies in the final state is exactly
the same as in the initial state ($E_{1f}=E_{2i}$, $E_{2f}=E_{1i}$), therefore,
we, in fact, have  imposed an additional constraint.
 For classical processes this event has zero phase volume (zero probability).

This work is  supported by the Australian Research
Council. I am grateful to J. Berengut, M.Yu. Kuchiev and V.G. Zelevinsky
  for useful discussions.

\end{document}